# Observation of Landau levels of Dirac fermions in graphite

Guohong Li and Eva Y. Andrei

*Department of Physics and Astronomy, Rutgers University, Piscataway, New Jersey 08854, USA*

**The low energy electronic excitations in single layer and bilayer graphite (graphene[1-2]) resemble quantum-relativistic particles also known as Dirac Fermions (DF) [3-5]. They possess an internal degree of freedom, chirality, that leads to unusual Landau Level (LL) energy sequences in a magnetic field[6-14] and profoundly alters the magneto-transport properties. One of the consequences is an anomalous Quantum-Hall effect, recently detected in both single layer and bi-layer graphene[7, 8,11]. However the underlying cause, the unusual LL sequence, was never observed. Here we report the direct observation of LL of DF by means of low temperature Scanning-Tunnelling-Spectroscopy (STS) on the surface of graphite in fields up to 12 Tesla. We find evidence of coexistence of massless and massive DF, and identify the zero-energy LL which is a unique consequence of their quantum-relativistic nature. Surprisingly these strictly two-dimensional properties emerge even on bulk graphite in samples where the interlayer coupling is weak.**

The dynamics of electrons inside a material is usually described within the framework of non-relativistic quantum mechanics. For parabolic energy bands the energy-momentum dispersion $E = E_{\pm} \pm \hbar^2 k^2 / 2m^*$, resembles that of non-relativistic free particles with an effective mass $m^*$ arising from interactions with the lattice. Here the energy is measured relative to the bottom of the conduction band, $E_{+} = E_C$, for electron-like (+) particles, or to the top of the valence band, $E_{-} = E_V$, for hole-like (-) particles. In the presence of a magnetic field, $B$, the energy for motion perpendicular to the field is quantized in a series of equally spaced LL:

$$E_n = E_\pm \pm \hbar\omega_c(n+1/2), \qquad n = 0, \ 1, \ 2,... \qquad (1)$$

where $\hbar$ is Planck's constant, $\omega_c = eB/m^*$ the cyclotron frequency and $e$ the electron charge.

This standard model fails radically when applied to electrons in graphene - a two-dimensional layer of carbon atoms tightly packed into a honeycomb structure[1,2] consisting of two distinct triangular Bravais sublattices. This arrangement has non-trivial consequences: it endows the electronic states with internal degrees of freedom, isospin, leading to well defined chirality and produces a highly diminished Fermi "surface" consisting of two inequivalent "Dirac Points" (DP), where the valence and conduction bands touch. The Hamiltonian describing the low energy excitation is that of a (2+1) relativistic quantum system[4,5] described by the Dirac equation for particles with zero mass and spin 1/2. The solutions, consisting of particle-antiparticle pairs with opposite chiralities, resemble relativistic massless DF in all except that they move at the Fermi velocity $v_F \sim c/300$ instead of the speed of light, $c$. Thus, the low energy excitation spectrum measured with respect to the DP is linear in momentum, $\hbar k$, with $E(k) = \pm v_F \hbar k$.

The relativistic nature of massless DF is revealed through an unusual LL energy sequence which consists of a field independent state at zero energy followed by a sequence of levels with square root dependence in both field and level index, instead of the usual linear dependence:

$$E_n = \text{sgn}(n)\sqrt{2e\hbar v_F^2 |n| B}, \qquad n = ... -2 \ -1, \ 0, \ 1, \ 2, \ ... \qquad (2)$$

Here the energy is measured relative to the DP energy, $E_D$. The unusual appearance of a zero energy level for $n = 0$, is a direct consequence of chirality. This level has the same degeneracy as the $n \neq 0$ levels, but it is the only one independent of field.

When two graphene layers stack to form a bilayer, the interlayer coupling leads to the appearance of band mass but it does not open a gap at the DP. Here the quasiparticles are still chiral and the LL spectrum takes the form [11-13] $E_n = \pm \hbar \omega_c \sqrt{n(n-1)}, \quad n = 0,1,2...$ This sequence is linear in field, similar to the standard case, but it contains an additional zero energy level which is independent of field. Here, as in the case for massless DF, the zero energy level is a consequence of chirality, but its degeneracy is now double that of the $n \neq 0$ levels. In order to facilitate the data analysis we will use an alternative form:

$$E_n = sgn(n)\hbar \omega_c \sqrt{|n|(|n|+1)}, \qquad n = ...-2, \ -1, \ 0, \ 1, \ 2... \qquad (3)$$

This form gives the same LL spectrum including the double degeneracy of the zero energy level. Indirect evidence of a zero energy LL was recently obtained through the appearance of QHE anomalies observed in single and bi-layer graphene [7,8,11] as well as in Highly Oriented Pyrolitic Graphite (HOPG)[15,16]. However, in order to observe the LL directly, a more specialized technique is required. STS allows selecting specific energy levels by varying the voltage bias between sample and tip and the tunnelling current gives direct access to the density of states. Obtaining tunnelling spectra on graphene films is challenging because the films are quite small and locating them in the field of view of a low temperature Scanning Tunnelling Microscope can be daunting. Furthermore most substrate materials interfere significantly with the intrinsic energy levels. Surprisingly we found that the LL of both massless and massive DF can be accessed on the surface of bulk HOPG samples, suggesting decoupling of the surface states from the bulk. This is in contrast to STS results on Kish graphite where the LL spectra are consistent with bulk[17].

Figure 1a shows tunnelling spectra measured on the surface of HOPG at 4.4 K in magnetic fields up to 12 T. Similar spectra on HOPG graphite were recently reported by another group[18]. We note that all peaks, except the two labelled *A* and *Z*, shift away from the origin with increasing field. Peak *A* is independent of field, while peak *Z* exhibits weak non-monotonic field dependence consistent with that previously associated with the $n = 0$, -1 bulk Landau levels of graphite[17]. Analyzing the field dependence of the spectra in Figure 1a, we find two families of peaks: one with square-root dependence and the other linear. Two peaks in each family are highlighted with stars and open circles for the square-root and linear sequences respectively. The square-root field dependence of the former becomes evident by plotting the peak positions against $B^{1/2}$ in Figure 1b. We note that the positive energy sequence has positive slope as expected of electron-like excitations while the negative slope of the negative energy sequence indicates hole-like behaviour. The field dependence of the levels in both sequences extrapolates to the same zero-field value, $16.8 \pm 3$ meV, implying that, within experimental error, the conduction and valence bands touch and that the intersection point coincides with the field-independent peak A. Comparing to equation (2) suggests that the square-root sequence corresponds to the LL of massless DF, in which case one needs to include peak *A* and identify its energy with the DP, $E_D$. A deviation of the DP from zero, also observed in ARPES[19], is often attributed to substrate charging. The positive value of $E_D$ measured here implies a hole doped sample.

We now proceed to analyze the dependence on LL index, *n*. From equation (2) the LL energy of massless DF scales according to $E/B^{1/2}$, where *E* is referred to the DP. We therefore re-plot in Figure 2 the spectra against the scaled energy $E/B^{1/2}$ with the energy origin shifted to the DP, $E-E_D \rightarrow E$. It is now straightforward to identify a family of peaks that align on the same value of scaled energy. All the aligned peaks are then indexed starting with the $n = 0$ level (peak *A*) at the origin. The unaligned peaks are presently left out. When plotting in Figure 2c the scaled energy of the aligned peaks against $|n|^{1/2}$, the square root dependence on level index becomes evident.

The family of peaks with linear field dependence ( Figure 1c) is considered next. As before the sequence with positive (negative) energy is attributed to electron-like (hole-like) particles. Both sequences extrapolate to the same energy at zero field, indicating that the conduction and valence bands touch. Notably this energy coincides with the energy $E_D$ found in the extrapolation of the square-root sequence. The linear field dependence implies finite mass[20] but whether it represents standard particles (equation (1)) or chiral ones (equation (3)), can only be determined by analyzing the dependence on LL index. In Figure 3a we plot the low field spectra against the scaled variable $E/B$. This scaling leads to the alignment of all peaks associated with massive particles, but those associated with massless particles (square-root sequence) become unaligned. Note that peak Z does not scale with either $B$ or $B^{1/2}$ so it is not included in either sequence. In order to assign a LL index to these peaks we need to establish whether peak A which is pinned to the Dirac point, should be considered part of this family. If we assume standard massive fermions, then peak A should be left out. In this case the first peaks in the sequence (marked with open circles) have to be labelled $n = -1$ and $n = +1$ for the negative and positive energies, respectively. Any other choice would lead to $E_v \neq E_c \neq E_D$. But if this assignment is correct then the two $n = 0$ levels, marked with arrows in Figure 3b, are inexplicably missing. If we assume the spectrum corresponds to massive DF, then peak A should be included as the $n = 0$ level and the other aligned peaks are labelled in ascending order as shown in Figure 3a. Plotting the scaled energy for this sequence against $(|n|(|n|+1))^{1/2}$ in Figure 3c we now find agreement with equation (3) in support of the interpretation in terms of massive DF. From the slope of this data we obtain an effective mass of $m^* = (0.028\pm0.003)\, m_e$ for both electrons and holes. This value is smaller than that of the parabolic band in bulk graphite ($m^* = 0.069 m_e$) obtained with ARPES [19].

The data for all resolved Landau levels, shown in Figure 4, can be classified into three groups: massless DF (Fig. 4a), massive DF (Fig. 4b), and unidentified peaks (Fig. 4c). A classification of the last group requires detailed band structure calculations which will be reported elsewhere. Here we only wish to point out that for the massive electrons and holes in this category the condition $E_v = E_c = E_D$ is generally not valid.

The observation of STS spectra corresponding to massless and massive DF together with the fact that their energies extrapolate to the same DP at zero field is a strong indication of purely two dimensional (2d) quasiparticles. This is expected in samples of few-layer graphene [21] but is quite surprising for bulk graphite because there, although both types of DF are present[3,19], the corresponding DP are shifted with respect to each other by ~ 50meV. Moreover the spectral weight of the 2d quasiparticles in graphite is negligible compared to that of bulk excitations along the H-K direction and therefore they should not be detectable with a technique, such as STS, which is not momentum selective[17]. The findings reported here strongly suggest that for HOPG samples only a few top layers of the graphite contribute to the observed electronic states.

To understand the origin of the observed 2d behaviour we first consider the tip induced charging of the graphite surface. The proximity of the tip causes an excess surface charge density[22] of $n \approx 5 \times 10^{12} (V_b / d) cm^{-2}$, where $V_b$ is the bias voltage in Volts and $d$ the tip-sample distance in nanometers. For our experiment (d~ 8nm) this causes a shift in Fermi energy (< 1meV) which is too small to affect the analysis. This was confirmed by repeating the experiments for several values of the tip-sample distance and finding that the spectra are independent of tip distance for all set currents below 100 pA (the data shown here were taken at 25 pA). Only when the currents become much larger (smaller tip-sample distance) do we observe a small shift in Fermi energy together with an increase of up to 10% in $m^*$ of the massive DF. This finding demonstrates that the observed 2d behaviour at large sample-tip distance is not a simple charging effect. Indeed if this were the case, the 2d quasiparticles would be observable in the STS spectra of all types of graphite, but their absence in Kish graphite[17] suggests that material structure plays an important role. Thus it was recently proposed that the presence of stacking faults, or when the stacking is rhombohedral rather than Bernal, the surface layers of graphite become well decoupled from the bulk giving rise to both massless and massive quasiparticles[23]. This is consistent with our observations. However more detailed theoretical modelling is needed to understand the STS spectra and the 2d nature of the quasiparticles on the surface of graphite.

**Methods**

Experiments were carried out on a home-built low temperature high magnetic field scanning tunneling microscope (STM) recently developed in our group. The controller was a commercial unit SPM 1000 controller from RHK Technologies. The samples were HOPG cleaved in air and the STM tips were obtained from mechanically cut Pt-Ir wire. The tip-sample distance was set by a tunnelling current of 25 pA with sample bias voltage of 300 mV otherwise specified. We use the small set-current to minimize the effect of tip on the tunnelling spectra. The data shown here represent the average of 256 spectra measured over an area of $50 \times 50$ nm$^2$. The dI/dV measurements were carried out with a lock-in technique using a 450 Hz ac modulation of the bias voltage with amplitude of 5 mV. The magnetic field was applied perpendicular to the sample surface with a superconducting magnet working in persistent mode.

**ACKNOWLEDGEMENTS**

We thank A. H. Castro Neto, P. Coleman, J.C. Davis, X. Du, A. Ermakov, A. Geim, F. Guinea, E. Garfunkel, P. Kim and I. Skachko for useful discussions. Work supported by NSF-DMR-0456473 and by DOE DE-FG02-99ER45742.

**LIST OF REFERENCES**

1. Novoselov, K. S. *et al.* Electric field effect in atomically thin carbon films. *Science* **306**, 666-669 (2004).

2. Novoselov, K. S. *et al*. Two dimensional atomic crystals. *Proc. Natl. Acad. Sci.USA* **102**, 10451-10453 (2005).

3. Wallace, P. R. The Band theory of Graphite. *Phys. Rev.* **71**, 622-634 (1947).

4. Semenoff, G. W. Condensed matter simulation of a three-dimensional anomaly. *Phys. Rev. Lett.* **53**, 2449-2452 (1984).

5. Haldane, F. D. M. Model for a quantum Hall effect without Landau levels: condensed matter realization of the "parity anomaly". *Phys. Rev. Lett.* **61**, 2015-2018 (1988).

6. Zheng, Y. & Ando T. Hall conductivity of a two-dimensional graphite system. Phys. Rev. B **65**, 245420 (2002).

7. Novoselov, K. S. *et al.* Two-dimensional gas of massless Dirac fermions in graphene. *Nature* **438**, 197-200 (2005).

8. Zhang, Y., Tan, Y. W., Stormer, H. L. & Kim, P. Experimental observation of the quantum Hall effect and Berry's phase in graphene. *Nature* **438**, 201-204 (2005).

9. Gusynin, V. P., & Sharapov, S. G. Unconventional integer quantum Hall effect in graphene. Phys. Rev. Lett. **95**, 146801 (2005).

10. Castro Neto A. H., Guinea F. and Peres N. M. R. Edge and surface states in the quantum Hall effect in graphene. Phys. Rev. B **73**, 205408 (2006).

11. Novoselov, K. S. *et al.* Unconventional quantum Hall effect and Berry's phase of $2\pi$ in bilayer graphene. *Nature Physics* **2**, 177-180 (2006).

12. Nilsson, J., Castro Neto, A. H., Peres, N. M. R. and Guinea F. Electron-electron interactions and the phase diagram of a graphene bilayer. Phys. Rev. B **73**, 214418 (2006).

13. MaCann, E. & Fal'ko, V. I. Zhang, Y. Landau-level degeneracy and quantum Hall effect in a graphite bilayer. *Phy. Rev. Lett.* **96**, 086805 (2006).

14. Peres, N. M. R., Guinea, F. & Castro Neto A. H. Electronic properties of disordered two-dimensional carbon. Phys. Rev. B **73**,125411 (2006).

15. Kopelevich Y. *et al.* Reentrant metallic behavior of graphite in the quantum limit. *Phys. Rev. Lett.* **90**, 156402 (2003).

16. Luk'yanchuk I.A. & Kopelevich Y. Dirac and normal fermions in graphite and graphene: implications of the quantum Hall effect. *Phys. Rev. Lett.* **97**, 256801 (2006).

17. Matsui T. *et al.* STS observations of Landau levels at graphite surfaces. *Phys. Rev. Lett.* **94**, 226403 (2005).

18. Niimi Y., Kambara H., Matsui T., Yoshioka D. & Fukuyama H. Real-space imaging of alternate localization and extention of quasi-two-dimensional electronic states at graphite surfaces in magnetic fields. *Phys. Rev. Lett.* **97**, 236804 (2006).

19. Zhou, S.Y. *et al.* A. First direct observation of Dirac fermions in graphite. *Nature Physics* **2**, 595-599 (2006).

20. Dresselhaus, G. Graphite Landau levels in the presence of trigonal warping. *Phys. Rev. B* **10**, 3602-3609 (1974).

21. Guinea, F., Castro Neto, A. H. & Peres, N. M. R. Electronic states and Landau levels in graphene stacks. *Phys. Rev. B* **73**, 245426 (2006).

22. Morozov S.V. *et al.* Two-dimensional electron and hole gases at the surface of graphite. *Phys. Rev. B* **72**, 201401 (2005).

23. Guinea, F., private communications.

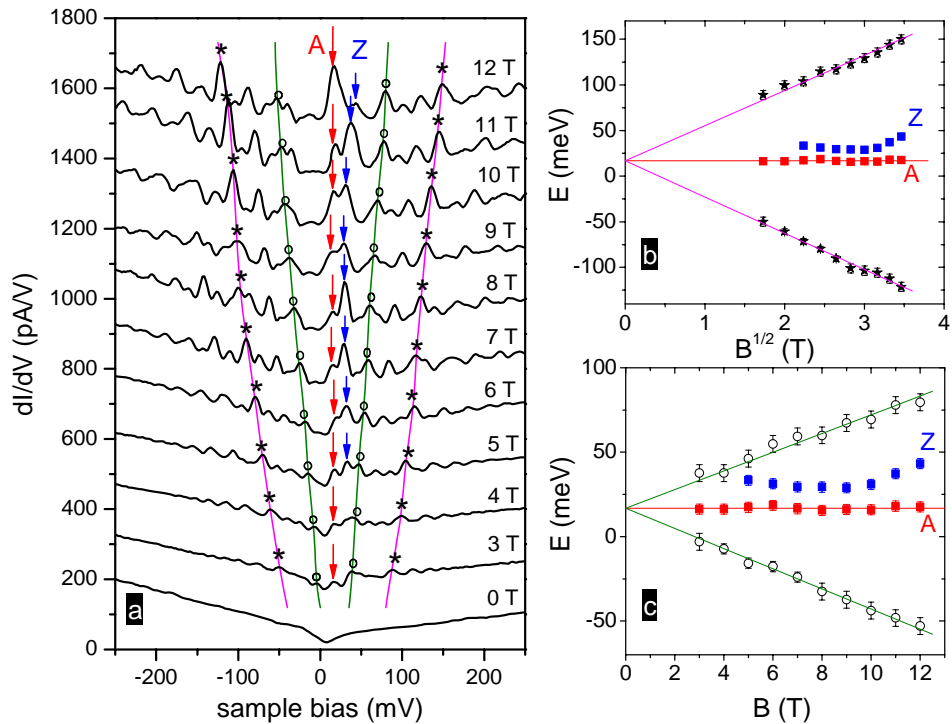

**Figure 1 a, Magnetic field dependence of tunnelling spectra on a graphite surface at 4.4 K.** All peaks, excepting the two labelled A and Z, show strong field dependence shifting away from zero bias with increasing field. The spectra are shifted vertically by 150pA/V for clarity. **b,** Landau levels marked with stars in Fig. 1a show square-root field dependence and extrapolate to the energy of peak A at zero field. Also shown is the field dependence of peaks A and Z. **c**, Landau levels marked with circles in Fig. 1a exhibit linear field dependence and extrapolate to the energy of peak A at zero field.

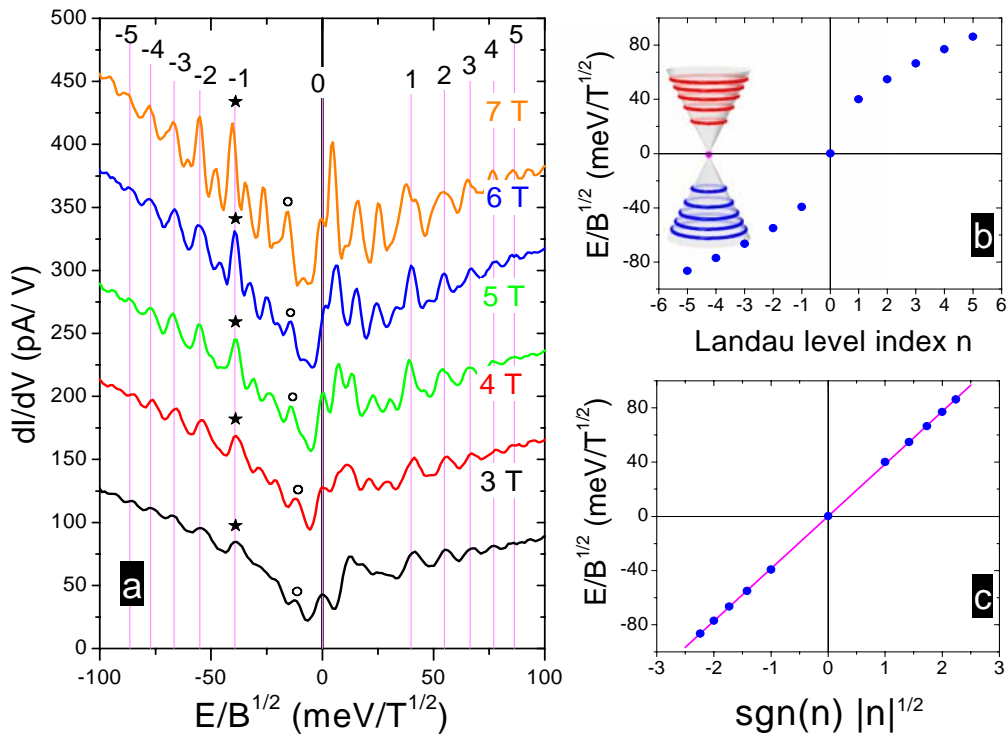

**Figure 2 Landau levels of massless Dirac fermions. a**, Tunnelling spectra plotted against the reduced energy $E/B^{1/2}$. The energy origin is shifted to peak A in Fig.1. The peaks marked with stars are aligned after scaling, while those marked with circles are not. All aligned peaks are labelled with sequential Landau level indices. The spectra are shifted vertically by 70 pA/V for clarity. **b**, Scaled energy of the aligned peaks plotted against Landau level index. Inset: Landau levels of massless Dirac fermions superposed on the Dirac cone. **c**, Scaled energy levels are linear in $\mathrm{sgn}(n)|n|^{1/2}$, as expected for massless Dirac fermions. From a comparison of the slope of the linear fit with equation (2) we obtain the Fermi velocity: $v_F = (1.07\pm 0.05)\times 10^6$ m/s, consistent with that obtained from transport measurements on graphene[7,8] and ARPES on graphite[19].

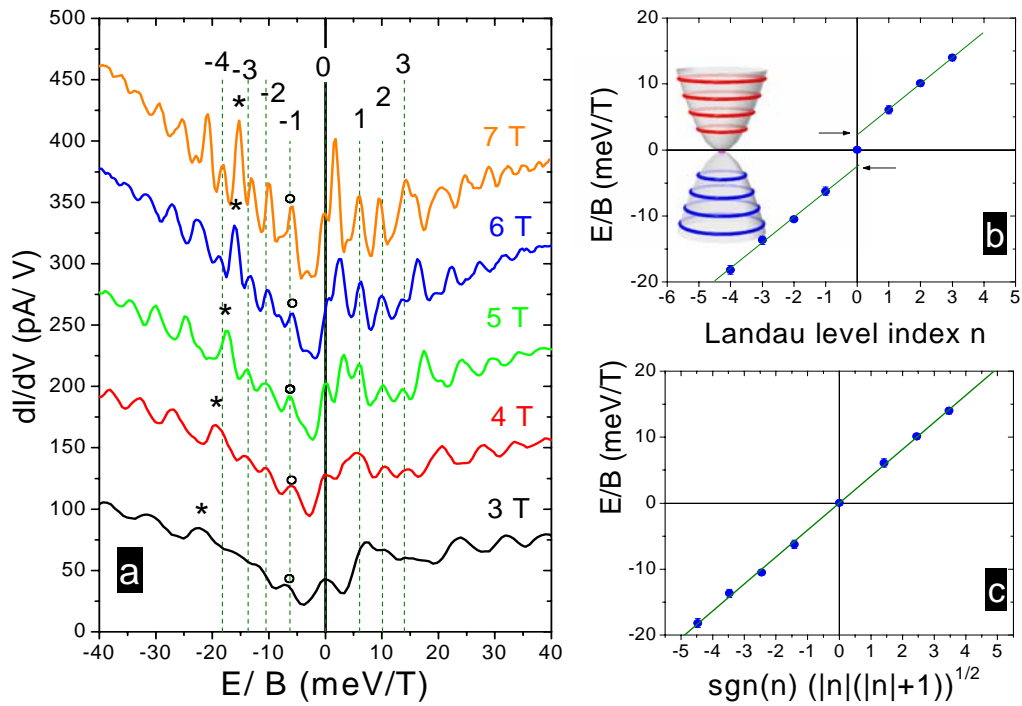

**Figure 3 Landau levels of massive Dirac fermions. a**, Tunnelling spectra plotted against the reduced energy E/ B. The energy origin was shifted to peak A in Fig.1. The peaks marked with circles are aligned after scaling, while those marked with stars are not. All aligned peaks are labelled with sequential Landau level indices. The spectra are shifted vertically by 70 pA/V for clarity. **b**, Scaled energy plotted against LL index. Solid lines represent the LL spectrum expected for normal massive fermions. The arrows show two missing n=0 levels. Inset: Landau level energy sequence of massive DF superposed on the zero-field dispersion. **c**, The scaled energy is shown to be linear in sgn(n)(|n|(|n|+1)$^{1/2}$, as expected for massive chiral fermions. The solid line is a linear fit with effective mass m$^*$ = 0.028m$_e$.

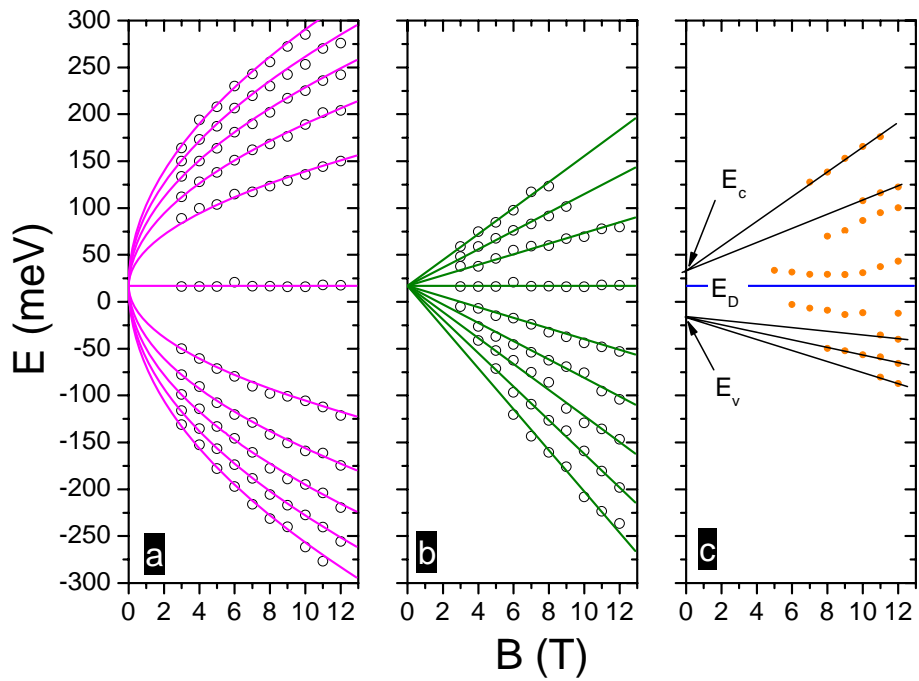

**Figure 4 Classification of Landau levels in magnetic field up 12 T. a,** Massless Dirac fermions. Solid lines represent fits of the spectra in Fig.2. with equation (2) **b**, Massive Dirac fermions. Solid lines represent fits of the spectra in Fig.3. with equation (3) **c**, Unidentified levels. The solid lines show that the Landau levels do not coalesce into a single energy at zero field.